\documentstyle[prl,aps,floats,epsf]{revtex}

\begin{document}
\def\SNG{{\em Physical Review Style and Notation Guide}}
\def\LUG {{\em \LaTeX{} User's Guide \& Reference Manual}}
\def\btt#1{{\tt$\backslash$\string#1}}%
\def\REVTeX{REV\TeX}
\def\AmS{{\protect\the\textfont2
        A\kern-.1667em\lower.5ex\hbox{M}\kern-.125emS}}
\def\AmSLaTeX{\AmS-\LaTeX}
\def\BibTeX{\rm B{\sc ib}\TeX}

\twocolumn[\hsize\textwidth\columnwidth\hsize\csname@twocolumnfalse%
\endcsname


\title{Composite fermions in the half-filled lowest Landau level:
       a macroscopic justification}


\author{F.\ Evers}


\address{Universit\"at Karlsruhe, TKM, D-76128 Karlsruhe, Germany} 

\date{\today}

\maketitle

\begin{abstract}
 An effective Hamiltonian for spinless electrons in the lowest Landau level
(LLL) close to half filling is derived.
As opposed to the standard treatment in Chern-Simons theories (CS)
we first project to the LLL and only then apply a CS-transformation
on the Hamiltonian. 
The transformed field operators act in the lowest Landau level only
{\it and} have fermionic commutation relations for small wavenumbers
ignoring gauge field fluctuations. When acting on the Hamiltonian at half filling
the {\it gauge transformation removes the monopole term in the
interaction and does not eliminate the magnetic field.} \\\mbox{ }
\end{abstract}
]


Ten years ago J. Jain made the discovery that the fractional quantum Hall effect
of interacting electrons may be understood in terms of the integer
quantum Hall effect of Composite Fermions (CF) \cite{Jain89}. 
The key property of this species is that they are fermions, their mutual
interaction is small and their 
interaction with the external magnetic field is reduced as compared to
electrons. One way the latter is frequently visualized is by
saying that CFs have charge $-e$ and interact with an effective
magnetic field that equals the external one reduced by two
flux quanta per particle:
$B_{\rm eff} = B_{\rm ex}-2\phi_0 n$.

By now the experimental evidence for the existence of CFs is
overwhelming \cite{Willett97,Smet98} and generally undisputed.
In particular, the existence of a well defined Fermi surface near
filling fraction 1/2, where the effective magnetic field vanishes has been
established. 
In contrast, the theoretical understanding, especially near
half filling which is what we focus on in the present letter,
is still somewhat incomplete. Important insights have come from numerical work,
in particular from the analysis of trial wave functions for finite
systems \cite{Jain98,Read94}. However, despite of the fact that a
number of attempts have been made to develop effective theories
\cite{Halperin93,Haldane97,Shankar97,Read98,Lee98,Stern98}
the efforts have been only partially successful.

Let us briefly review two of these approaches. The most prominent one
has been developed by Halperin,
Lee and Read (HLR)\cite{Halperin93}. Originally, the most intriguing
feature of the theory was the prediction of a well defined Fermi
surface for the CF's at $\nu=1/2$.
It manifested itself in the RPA-density response 
that explained an anomaly in surface acoustic wave
experiments \cite{Willett97}. However, as has been
pointed out e.g. by Simon \cite{Simon96}, the theory is not capable of
properly accounting for zero order physics in high magnetic fields
unless these effects are put in by hand. For example, a well known
problem is that up to now it has not been possible to demonstrate
how it happens within this formalism that the bare mass $m_{\rm b}$
describing the energy dispersion of free electrons is
replaced by an effective mass $m^*$ describing
the dispersion of the CFs. Since the latter is due electron interactions only,
$m^*$ must be independent of $m_{\rm b}$.

The difficulty lies in the very nature of the approach itself:
As the first step HLR perform a singular gauge
transformation (Chern-Simons transformation). It cancels the
external magnetic field at half filling on average and  
virtually {\it all} of the physics related to the actual existence of a strong
magnetic field must therefore come from gauge field fluctuations.
Apparently, already RPA theory can give proper account of the structure
of the density response, but many other features, like the effective
mass, have not been properly restored from the fluctuations up to now. 

An interesting recent attempt in this direction has been made by
Shankar and Murthy \cite{Shankar97}. However, success in their
theory depends on  a 
parameter $\cal Q$ that is not fixed within the theory itself.  For a choice
${\cal Q}=k_{\rm F}$ the effective mass is indeed independent of
$m_{\rm b}$ at the expense of a density response that is
incompressible in disagreement with what one expects for the Fermi
liquid \cite{Halperin98}. Stern {\it et. al.}
advocate another choice ${\cal Q}\ll k_{\rm F}$ which
reproduces the expected density response and also an effective
Hamiltonian with a desirable structure but now the mass is on the
wrong scale, again\cite{Stern98}.

The second type of approach considers the fermions as bosons at
integer filling carrying one flux quantum \cite{Haldane97,Read98}.
The theory deals with the Hamiltonian projected to the LLL
and discards the idea of a singular gauge transformation, altogether.
Generally, the Hamiltonian of a projected theory is formulated in
terms of projected field operators that are not easily dealt with
since they do not anticommute.
The method for handling the projected field operators chosen in
this type of theory is to replace the projected operators by
the unprojected counterpart and complement the Hamiltonian
with a constraint.
Then the main problem is to deal with the constraint.
Read has calculated the density response and his result is 
structurally identical with the HLR response. 

As it stands, the main disadvantage of this theory is that
it has been formulated for bosons at filling fraction unity
and the relation to electrons at half filling is not entirely obvious.
Moreover, it has not been extended to filling fractions
other than $\nu=1$, yet.

In the present paper we combine ideas underlying the earlier theories.
By projecting to the LLL we first implement
the basic physics of strong magnetic fields: we eliminate the bare
kinetic energy in the Hamiltonian and introduce the guiding center
coordinate
as a dynamical variable in the interaction term.
Then we carry out a Chern-Simons transformation which introduces the CFs.
The resulting theory reproduces correct results from the earlier
CS-approaches in a straightforward manner but is free from their problems.  
The coupling to disorder is different however and might lead to
experimentally observable differences from the standard theory. 

%

In contrast to all previous CS-theories our motivation for
applying the CS-transformation focusses on the fermionic
character of the resulting quasiparticles:
our ultimate goal is to formulate a theory in terms of
quasiparticles that live in the LLL only {\it and} have
fermionic statistics, at least in the small wavenumber limit.
In principle, other choices for the statistics are possible, but
fermions are preferable since they obey the Pauli principle which automatically
favors configurations with low interaction energy independent of
details of the (repulsive) interaction. 
In order to see how to construct the quasiparticles,
consider the anticommutator 
\begin{equation}
\label{eq:3}
[\psi_{\cal P}({\bf x}),\psi_{\cal P}^\dagger({\bf x'})]_{+}
= \frac{1}{2\pi}e^{-i/2\ {\bf x}\wedge{\bf x'}}
e^{-|{\bf x}-{\bf x'}|^2/4}
\end{equation}
of the projected field operator $\psi_{\cal P}({\bf x})$ 
defined below. 
(Throughout the paper we use the symmetric gauge, choose units
such that $\ell=\sqrt{\hbar c/eB}=1$ and ${\bf x}\wedge{\bf x'}=x_1 x_2'-x_2 x_1'$.) 
The argument of the complex exponential is the phase an electron
picks up when moving around the triangle $0,{\bf x},{\bf x'}$.
One can introduce new particles with the gauge transformation
\begin{equation}
\label{eq:4}
    \chi({\bf x}) = \exp\!\!\left(\!-i\tilde\phi \!\int\!\! d{\bf r}
                    \arg({\bf x}-{\bf r})\ 
                    \psi_{\cal P}^\dagger({\bf r})\psi_{\cal P}({\bf r})\!\right)
		    \psi_{\cal P}({\bf x})
\end{equation}
where $\arg({\bf r})$ denotes the angle between the argument and the
x-axis and $\tilde\phi$ is the number of flux quanta that are
attached. The Pauli principle requires it to be an even number,
e.~g. $\tilde\phi=2$ is the proper choice at half filling. 
The key observation is that these particles $\chi$ obey the  anticommutator
\begin{equation}
\label{eq:5}
[\chi({\bf x}),\chi({\bf x'})^\dagger]_{+}
= \frac{1}{2\pi} e^{-|{\bf x}-{\bf x'}|^2/4}
\end{equation}
without a complex phase factor
in an approximation where the effective transverse gauge field (operator)
\begin{equation}
\label{eq:6}
{\cal A}({\bf x}) = \tilde\phi \nabla\!\int\! d{\bf r} \arg({\bf x}-{\bf r})
	     \chi^\dagger({\bf r})\chi({\bf r})
	   - A_{\bf ex}({\bf x})
\end{equation}
is neglected \cite{Evers99a}. This corresponds to mean field approximation at half
filling. As long as the mean particle distance is large as compared to
the width $2\ell$ of the Gaussian they can be considered
fermions. Note, that in the literature \cite{Halperin93,Fradkin91} the analog
object for the standard gauge transformation is frequently called
${\bf a}({\bf x})$ \cite{rem1}.
We have introduced a different symbol in order to stress
that ${\bf a}({\bf x})$ is related to fluctuations of the electron density
whereas ${\cal A}({\bf x})$ relates to the quasiparticle density
$\nabla\wedge{\cal A}({\bf x}) = 2\pi\tilde
\phi\chi^\dagger({\bf x})\chi({\bf x})-B_{\rm ex}$.
These are different quantities in the present theory and we will come
back to this crucial point, below. 

Before we discuss some of the properties of the new CFs we derive the full model: 
The Hamiltonian underlying the problem is (${\bf A}_{\bf ex}=B/2(-y,x)$, $V$:
Volume, $e/c=1$)
\begin{eqnarray}
\label{eq:6.1}
    H &=& \int \!d{\bf x} \psi^\dagger({\bf x})\ \frac{1}{2m_{\rm b}}
                        (\nabla/i - {\bf A}_{\bf ex})^2 \ \psi({\bf
                        x}) \\\nonumber
        && + \frac{1}{V}\sum_{{\bf q}} v({\bf q}) \rho({\bf q}) \rho(-{\bf q}) 
\end{eqnarray}
$\rho({\bf q}) = \int\!d{{\bf x}} \psi^\dagger({\bf x}) e^{i{\bf qx}} \psi({\bf x})$. 
The projection to the LLL is accomplished by making use of a
standard formalism \cite{Weidenmueller87}:
we introduce single particle eigenstates $|jm\rangle$  with Landau
level index $j$ and an inner quantum number $m$ that is related to
angular momentum. New field operators $c_{j,m} = \int\!d{\bf x}
\ \psi({\bf x}) \ \langle jm|{\bf x}\rangle$ may be defined and in
terms of these the projected electron field operators are
$\psi_{\cal P}({\bf x}) = \sum_{m} c_{0,m} \langle{\bf x}|0m\rangle$.
The projected Hamiltonian reads
\begin{equation}
\label{eq:7}
    H = \omega_{\rm c}/2\  \rho_{\cal P}({\bf q}=0)
    + \frac{1}{2V}\sum_{{\bf q}\not= 0} v_0({\bf q}) 
                     \rho_{\cal P}({\bf q}) \rho_{\cal P}(-{\bf q}) 
\end{equation}
with the effective interaction $v_0(q)=v(q)\exp(-q^2/2)$ and an operator 
$    \rho_{\cal P}({\bf q}) =  \int\!d{\bf x} \ \psi_{\cal P}^\dagger({\bf x})\ 
     e^{i{\bf qR}} \ \psi_{\cal P}({\bf x}) $
representing the electron density. 
The important feature is that the position operator in the
plane wave factor $\exp i{\bf qx}$ of the ordinary expression for the
density operator (below eq.(\ref{eq:6.1})) has been replaced by the
guiding center coordinate
${\bf R} = {\bf x} + (\nabla/i - {\bf A}_{\bf ex})_{\rm t}$ with
${\bf x}_{\rm t}=(-y,x)$. 
It is sensitive to gauge transformations in which 
the external field is canceled in part by
the gauge field which leaves us with  $-{\cal A}$ instead of ${\bf
A}_{\bf ex}$: 
\begin{equation}
\label{eq:10}
    \rho_{\cal P}({\bf q}) =  \int\!d{\bf x} \ \chi^\dagger({\bf x})\ 
     e^{i{\bf q}{\bf x} + i{\bf q}\wedge(\nabla/i+{\cal A}({\bf x}))} \ \chi({\bf x}).
\end{equation}
Equations (\ref{eq:6}), (\ref{eq:7}) and (\ref{eq:10}) together with
fermionic commutators for the quasiparticles $\chi, \chi^\dagger$ constitute the
model for composite fermions that we consider here. 

Now, let us compare our theory to the standard ones and see what
differences arise due to the different gauge transformations used. 
1) In both theories the gauge transformation breaks the symmetry
   between wavenumber and position, that is the typical effect of
   magnetic fields and a prerequisit for introducing a mass term.
   The standard transformation achieves this by acting  
   on the kinetic term of the Hamiltonian eliminating the magnetic
   field on average. By contrast, our version affects the interaction term,
   only. We show below that on mean field level at $\nu=1/2$ 
   the gauge transformation actually removes the
   monopole term in the interaction and {\it not} the magnetic field.
   It leaves us with a dipole term $\chi^\dagger\nabla \chi
   \chi^\dagger\nabla \chi$.
2) The transformation of HLR maps fermions to fermions and does
   not modify the (anti)-commutators. The corresponding field
   operators couple different Landau levels and hence do not describe
   the experimentally confirmed, "real" CF.  On the
   other hand,  our projected field operators cannot be
   interpreted as proper fermions by themselves since they do not 
   anticommute. We use the gauge transformation in order to recombine
   these "improper particles" so as to form a composite object with
   fermionic commutator, at least approximately. As opposed to
   ordinary metals, in our case the screening
   cloud adds to the fermionic character of the quasiparticles and not
   only screens the charge. 
3) As already mentioned, the gauge field operator ${\cal A}$ is related
   to the quasiparticle density $\chi^\dagger\chi$
   and not to the physical electron density $\rho_{\cal P}({\bf q})$.
   As can be seen from equation (\ref{eq:12}) these are quite
   different objects: at $\nu=1/2$ the physical
   electron density equals the transverse current mode of the
   quasiparticles and is not directly related to the quasiparticle
   density. This is to be contrasted with the standard theory where
   electron- and quasiparticle density are identical. 
   This difference might be relevant
   for the interpretation of experiments: electric fields produce 
   fluctuations in the electronic density that in turn induce an
   inhomogenous effective magnetic field seen by the CF. 
   In the standard theory the effective field is directly coupled
   to the induced charge density whereas in our setup it couples
   to the induced quasiparticle density.
   This opens up a possible mechanism that might be related to 
   quantitative discrepancies occuring in the theoretical description of
   several experiments \cite{Mirlin96,Evers99}.
   It is found that these experiments can be
   reasonably well understood within the CF picture,
   however with fitting parameters
   that point to larger scattering times than expected from
   estimates based on HLR theory by a factor of $2 \div 4$.
   A sufficiently weaker coupling to disorder than
   predicted by standard theory would explain the difference and
   appears at least possible within the framework suggested, here.

In the remainder of the letter we show that results
from the standard theory and its extension follow
in a straightforward manner within the scheme proposed.
We begin by deriving an expression valid at small
wavenumbers $q$ for
the density operator (\ref{eq:10}). Note that attention
should be paid to the gauge field: we decompose it into average
$\bar {\cal A}=(1-\tilde\phi\nu){\bf A_{\rm ex}}$
and fluctuations $\tilde {\cal A}$.
The derivative of the latter is of order $q^0$ since $\nabla\wedge\tilde{\cal
A}=2\pi\tilde\phi:\chi^\dagger\chi:$ with $:\chi^\dagger\chi: =
\chi^\dagger\chi - n$. We perform a double expansion and
keep terms up to linear order in $q$ and $\tilde{\cal A}$:
\begin{eqnarray}
\label{eq:11}
\rho_{\cal P}({\bf q})\!&=&  \!\int\!d{\bf x} e^{i{\bf q}{\bf x}}\ \chi^\dagger({\bf x}) 
                           \left(1 + i{\bf q}\wedge(\nabla/i+\bar{\cal
                           A})\right)\chi({\bf x})
			\nonumber\\ 
                      +\chi^\dagger({\bf x})(1 \!\!\!\!&&\!\!\!\!
                           +i{\bf q}\wedge(\nabla/i+\bar{\cal A}))\chi({\bf x}) 
                           \ i{\bf q}\wedge\tilde{\cal A}({\bf x}+{\bf q}_{\rm t}/2) + \ldots
                           \nonumber
\end{eqnarray}
After partial integration one obtains 
\begin{eqnarray}
\label{eq:12}
       \rho_{\cal P} &\approx& V\bar\rho\delta_{{\bf q},0} +
                          (1{-}\delta_{{\bf q},0})(1{-}\nu\tilde\phi)
                       \int\!d{\bf x} e^{i{\bf qx}} :\chi^\dagger({\bf x})\chi({\bf x}): \nonumber\\
                && + \int\!d{\bf x} e^{i{\bf qx}} i{\bf q}\wedge
                {\bf g}({\bf x})\nonumber \\
		&& - 2\pi\tilde\phi \int\!d{\bf x} e^{i{\bf qx}} i {\bf
		q}\wedge{\bf g}({\bf
                          x}):\chi^\dagger({\bf x})\chi({\bf x}):
\end{eqnarray}
where ${\bf g}=
1/2\ (\chi^\dagger(\nabla/i+\bar{\cal A}) \ \chi - (\nabla/i+\bar{\cal A}) \chi^\dagger \ 
\chi)$. The first three terms constitute  the
expression for the density operator as found and discussed previously by Shankar and
Murthy \cite{Shankar97}. It  can be interpreted consistently as
describing particles with a screened, effective
charge $e^*=e(1-\tilde\phi \nu)$ that couple to the external field
${\bf A}_{\rm ex}$. In particular, on the plateaus at filling fractions
$\nu=p/(p\tilde\phi+1)$ it is equal to the fractional charge of the
quasiparticles $e/(p\tilde\phi+1)$. At half integer filling fractions
$\nu=1/\tilde\phi$ the effective charge vanishes and the dipole moment
is the leading coupling to electric fields.

The fourth term is new  and 
related to K-invariance \cite{Stern98}:
under the transformation $\chi({\bf x})
\rightarrow \exp(i{\bf Kx})\chi({\bf x}).$ the density operator
acquires a phase factor $\exp i{\bf q}\wedge{\bf K}$ and the
Hamiltonian is unchanged. Its importance has been stressed
first by Halperin and Stern \cite{Halperin98}.
The additional term ensures the correct transformation properties
to leading order and would be absent on mean field level
($:\chi^\dagger\chi:\approx 0$).


Next we derive an effective Hamiltonian for our model. 
We restrict ourselves to half integer filling, here. Our strategy is to assume
that the quasiparticles $\chi({\bf x})$ form a well defined Fermi
sphere so that standard concepts of Fermi liquid theory can be
carried over. We begin with the usual phase space separation for the
Hamiltonian
\begin{eqnarray}
\label{eq:13}
H&=&\frac{1}{2V}\sum_{\bf q} v_0({\bf q}) \int \!d{\bf x}d{\bf x'}
\chi^\dagger({\bf x}) e^{i{\bf qR}+i{\bf q}\wedge{\tilde{\cal A}}({\bf x})} \chi({\bf x})\nonumber\\
&&\chi^\dagger({\bf x'}) e^{-i{\bf qR'}-i{\bf q}\wedge{\tilde{\cal A}}({\bf
x'})}\chi({\bf x'})
\end{eqnarray}
and subdivide it into regions with small and large momentum transfer.
In the small angle scattering region ${\bf q}\ll k_{\rm F}$
the expression (\ref{eq:12}) for the density operator holds
(with $\bar {\cal A}=0$) and we have
\begin{equation}
\label{eq:14}
H_{\rm dir} = \frac{1}{2}\sum_{{\bf q}\neq 0}^{\cal Q} v_0(q) \rho_{\cal P}({\bf
q})\rho_{\cal P}({\bf -q}).
\end{equation}
The momentum cutoff ${\cal Q}$ has been introduced in order to avoid
double counting. It satisfies the condition ${\cal Q}\ll k_{\rm F}$.
In a compressible system an appropriate choice for ${\cal Q}$ is the
screening wavenumber \cite{Evers98}.
Although we use a simplified low wavenumber expression
this contribution to the energy is still complicated since terms 
containing up to eight field operators are necessary in order to ensure
approximate K-invariance. On the other hand its contribution to
the energy is small. For Coulomb interactions it is of the order 
$o({\cal Q}\ell)$. 

For the exchange (or large angle scattering) contribution to 
the effective Hamiltonian a cancelation as for the direct
channel does not occur. Therefore, we treat the gauge field fluctuations in
the simplest approximation and ignore them.
This makes the exchange contribution well defined. It
can be found after shifting ${\bf q}\rightarrow {\bf q}+({\bf x}_1-{\bf x}_2)_{\rm
t}$:
\begin{eqnarray}
\label{eq:15}
H_{\rm ex} &=& \frac{1}{2}\sum_{\bf q}^{{\cal Q}} \sum_{{\bf k}_1,{\bf k}_2} 
             \ f({\bf k}_1 - {\bf k}_2)e^{i{\bf q}\wedge({\bf
             k}_1-{\bf k}_2)} \nonumber \\
             &&\chi^\dagger_{{\bf k}_1+{\bf q}/2} \chi_{{\bf k}_1-{\bf q}/2} 
             \chi^\dagger_{{\bf k}_2-{\bf q}/2} \chi_{{\bf k}_2+{\bf q}/2}
\end{eqnarray}
with
$             f({\bf k}) = 
              - \Theta(|{\bf k}|-{\cal Q})\ v_0({\bf k}).$ 
Equation (\ref{eq:15}) is a two-fermion interaction that is manifestly
K-invariant. It gives rise to a self consistency equation for the
self energy in Hartree Fock approximation 
\begin{equation}
\label{eq:17}
\Sigma_{\bf k} = - \frac{1}{V}\sum_{|{\bf p}-{\bf
k}|\geq{\cal Q}}v_0({\bf k} - {\bf p})
n({\Sigma_{\bf p}-\mu}) 
\end{equation}
where $n(\Sigma_{\bf k}-\mu)$ is the Fermi function. As already
mentioned the contribution of the direct term to $\Sigma_{\bf k}$ is
not easily treated systematically. Since we expect it to be small
we neglect it here.  From the self energy we
deduce an effective mass
\begin{equation}
\label{eq:18}
\frac{1}{m^*} = \frac{1}{k_{\rm F}}\frac{\partial\Sigma_{\bf
k}}{\partial |{\bf k}|} = 
\int_{|{\bf p}-{\bf k}|\geq{\cal Q}}\!  \frac{d\theta_{\bf kp}}{(2\pi)^2}
              v_0({\bf k}-{\bf p})\cos\theta_{{\bf k}{\bf p}}
\end{equation}
where $\theta_{\bf kp}$ denotes the angle between ${\bf k}$ and ${\bf
p}$ and $|{\bf k}| = |{\bf p}| = k_{\rm F}$. It is of the order of
$e^2\ell\ln({\cal Q}\ell)$ for Coulomb interaction. A similar result has been
reported recently by Read \cite{Read98}.

In Fermi liquid theory the standard way to treat a nonlocal
interaction $f({\bf k})$ is a mode 
decomposition. Thereby Fermi liquid parameters 
$f_0$ for density-density and $f_1$ for current-current modes
are introduced, where
$f_1=-2\pi/m^*$ in the present case. 
The resulting Hamiltonian reads
\begin{eqnarray}
\label{eq:19}
H &=&  \frac{e B_{\rm ex}}{m_{\rm b}c}\rho_{\cal P}({\bf q}=0) + \sum_{\bf k}
\frac{k^2}{2m^*}\ \chi^\dagger_{\bf k} \chi_{\bf k} \nonumber \\
  - && \frac{1}{2m^* n}\sum_{\bf q\neq0}^{\cal Q}
{\bf g}({\bf q}) {\bf g}({\bf -q}) +
\frac{1}{2}\sum_{{\bf q}\neq 0}^{\cal Q} v_0(q)\rho_{\cal P}({\bf q})\rho_{\cal P}({\bf -q}) \nonumber\\
+&&\frac{w_0}{2}\sum_{{\bf q}\neq0}^{\cal Q}  \rho_0({\bf q})\rho_0({\bf -q})
  +w_{\rm t} \sum_{{\bf q}\neq0}^{\cal Q}  {\bf q}\wedge{\bf g}({\bf
  q})\rho_0({\bf -q})
\end{eqnarray}
where $\rho_0=\chi^\dagger\chi$ and $\rho_{\cal P}$ as given in equation
(\ref{eq:12}). The coupling constants $w_0$ and $w_t$ equal $f_0$
on the current level of approximation. However they, as well
as $m^*$, might be strongly renormalized by gauge field fluctuations. 
A Hamiltonian similar in the leading terms  has been derived and
discussed by Stern {\it et. al.}, recently \cite{Stern98}.
Our result is different in two important respects:
1) The terms number five and six stem from an expansion of the
exponential in equation (\ref{eq:15}) and appear to be new. Of special interest
for us is the last term since it couples the transverse current (or electron
density) to the quasiparticle density. As outlined above this coupling
mediates the change in the effective magnetic field (or charge)
experienced by the CF when an electric field is applied.
2) Every particle has a
bare kinetic energy $eB_{\rm ex}/cm_{\rm b}$, where $m_{\rm b}$ is the
bare (or band) mass. The dispersive part of the energy is described
through an effective mass $m^*$ independent of $m_{\rm b}$ given in
equation (\ref{eq:18}).

We turn to our last issue, the density response. In RPA approximation
we find for the electron-density response function
\begin{equation}
\chi_{++} = \frac{1}{v(q)-\chi_d^*/n^2 + {\cal W} - 2i\omega/q^3k_{\rm F}n}
\end{equation}
where ${\cal W}=-w_{\rm t}^2/(w_0 +2\pi/m^*)$. This result is
structurally identical with the earlier result by
HLR \cite{Halperin93}.
The offdiagonal conductivity can be obtained with the notion that
$\omega_c({\bf x})$ acts like a chemical potential for a situation
with a filling fraction that is kept fixed, locally. This gives rise
to a relation between cyclotron current 
${\bf j}_{\rm c}=\hbar/2m_{\rm b} \nabla_{\rm t} \rho_{\cal P}$ and a
potential gradient from which we conclude that 
$\sigma_{\rm xy}=\nu$.


It is a pleasure to acknowledge valuable discussions with
W.~Apel, U. Gerland, T. Kopp, Mei-Rong Li, A.~D. Mirlin,
D.~G.Polyakov, A.~Rosch, J.~Wilke and
P. W\"olfle. Also, I am grateful to the Graduiertenkolleg {\it
Kollektive Ph\"anomene im Festk\"orper}
der Deutschen Forschungsgemeinschaft for financial support. \\



\begin{thebibliography}{XXXX99}
\bibitem{rem1}
In the standard transformation $\psi_{\cal P}$
in equation (\ref{eq:4}) is replaced by the unprojected field operator.
\end{thebibliography}
\end{document}